\documentclass[
	pra,
	aps,
	reprint,
	amsmath,amssymb,
	a4paper,
	superscriptaddress,
	10pt
]{revtex4-2}

\usepackage{graphicx}
\usepackage{physics}
\graphicspath{{fig/}}
\usepackage[dvipsnames,table,xcdraw]{xcolor}
\definecolor{rmpblue}{HTML}{2e3092}
\usepackage[
	colorlinks=true,
	citecolor=rmpblue,
	linkcolor=rmpblue,
	filecolor=magenta,
	urlcolor=rmpblue,
]{hyperref}
\usepackage[
	colorlinks=true
]{hyperref}

\usepackage{siunitx}
\newcommand{\um}{\SI{}{\micro\meter}}

\newcommand{\affilANU}{Nonlinear Physics Center, Research School of Physics, Australian National University, Canberra ACT 2601, Australia}

\newcommand{\affilEPFL}{Institute of Bioengineering, \'Ecole Polytechnique F\'ed\'erale de Lausanne (EPFL), Lausanne 1015, Switzerland}

\newcommand{\affilIT}{Department of Information Engineering, University of Brescia, Via Branze 38, 25123 Brescia, Italy}

\newcommand{\affilUS}{Aviation and Missile Center, U.S. Army CCDC, Redstone Arsenal, Alabama 35898-5000, USA}

\begin{document}
	
\title{Unconventional high-harmonic generation in resonant membrane metasurfaces}
\author{Pavel Tonkaev}
\altaffiliation{Contributed equally}
\affiliation{\affilANU}
\author{Felix Richter}
\altaffiliation{Contributed equally}
\affiliation{\affilEPFL}
\author{Ivan Toftul}
\affiliation{\affilANU}
\author{Maria Antonietta Vincenti}
\affiliation{\affilIT}
\author{Ivan Sinev}
\affiliation{\affilEPFL}
\author{Michael Scalora}
\affiliation{\affilUS}
\author{Hatice Altug}
\email{hatice.altug@epfl.ch}
\affiliation{\affilEPFL}
\author{Yuri Kivshar}
\email{yuri.kivshar@anu.edu.au}
\affiliation{\affilANU}
	
	
\begin{abstract}
High-harmonic generation (HHG) in solids has rapidly emerged as a promising platform for creating compact attosecond sources and probing ultrafast electron dynamics. Resonant metasurfaces are essential for enhancement of the otherwise small harmonic generation efficiency through local field enhancement and are essential to circumvent the need of phase matching constraints. Until now, the metasurface-enhanced HHG was believed to follow the conventional integer-power scaling laws that hold for non-resonant bulk HHG.  Here, we discover that highly resonant metasurfaces driven by quasi-bound states in the continuum break this principle, manifesting {\em non-integer intensity dependencies} of the generated harmonic powers. We show experimentally and theoretically that these unconventional nonlinearities arise from the high-Q resonances that generate local fields strong enough to substantially alter the contribution of higher order susceptibility tensors to the effective nonlinearities of the system. Our findings reveal how harmonic generation rooted in resonant field-driven modification of effective nonlinear susceptibilities can reshape our understanding of light–matter interaction at the nanoscale.

\end{abstract}
   
\maketitle

\section{Introduction}
High-harmonic generation (HHG) stands as a fundamental phenomenon at the intersection of strong-field physics, ultrafast science, and nonlinear optics. Initially observed in noble gases, where intense laser fields prompt electrons to re-collide with their parent ions emitting high-energy photons, HHG has revolutionised attosecond science by enabling the production of ultrashort pulses and the real-time probing of electron dynamics. This foundational mechanism has subsequently been extended to condensed matter physics, where HHG allows to uncover the intricate interplay between intense light fields and the electronic structure of solids. In this context, HHG in bulk crystals not only offers a pathway to more compact and scalable device platforms, but also serves as a sensitive probe of crystal symmetry, electronic topology, and ultrafast carrier dynamics.
A diverse array of materials underpins these advances, including non-centrosymmetric compounds such as ZnO  ~\cite{ghimire2011observation} and LiNbO$_3$~\cite{zhao2023efficient}, elemental semiconductors like Si~\cite{liu2018enhanced}, and III-V compound semiconductors like InP~\cite{shcherbakov2021generation} and AlGaAs~\cite{zalogina2023high}. These material platforms enable HHG applications ranging from compact EUV sources~\cite{Boukhaoui2025APLPhot} to spectroscopy of electronic band structures~\cite{Lanin2017Optica,TancogneDejean2017PRL}, with the orders of generated harmonics reaching record high values of several tens~\cite{Ghimire2011PRL,Apostolova2018DiamRelMat}.

With the advent of nanophotonics, metasurfaces have emerged as tools that offer unprecedented control over local electromagnetic fields and optical resonances which proved particularly useful for studying nonlinear phenomena and expanding their scope~\cite{vabishchevich2023nonlinear}. A key breakthrough has been the discovery of bound states in the continuum (BICs) and quasi-BICs, which enable ultrahigh-Q resonances by suppressing radiative losses through symmetry protection or mode superposition~\cite{koshelev2018asymmetric}. These modes have been exploited to enhance various nonlinear optical processes~\cite{carletti2018giant,koshelev2019nonlinear,kravtsov2020nonlinear,koshelev2020subwavelength,gao2024overcoming}, including HHG~\cite{carletti2019high,zograf2022high,zalogina2023high}, in ways unattainable in conventional bulk systems. For example, perovskite metasurfaces supporting quasi-BIC modes have demonstrated efficient fifth harmonic generation under picosecond pulsed excitation~\cite{tonkaev2023observation}. In contrast, symmetry-broken silicon metasurfaces pumped by picosecond pulses exhibit even harmonics (second and fourth) that deviate from expected power laws, pointing to resonant symmetry breaking and field enhancement effects \cite{tonkaev2024even}. Further pushing the capabilities of BIC-enabled metasurfaces, Zograf \textit{et al.} demonstrated odd harmonics up to the 11-th order generated from dielectric structures supporting BIC resonances~\cite{zograf2022high}.  The observation of HHG in other nanophotonic systems, including GST~\cite{korolev2024tunable}, CdT~\cite{long2023high}, ZnO~\cite{sivis2017tailored} and ENZ materials~\cite{yang2019high}, reflects a broader trend toward ultra-thin and strongly resonant platforms for high-field light–matter interaction. However, the interpretation of these phenomena is frequently complicated by multifaceted interactions with substrates, thermal effects, and material absorption.  

\begin{figure*}
    \centering
    \includegraphics[width=0.8\linewidth]{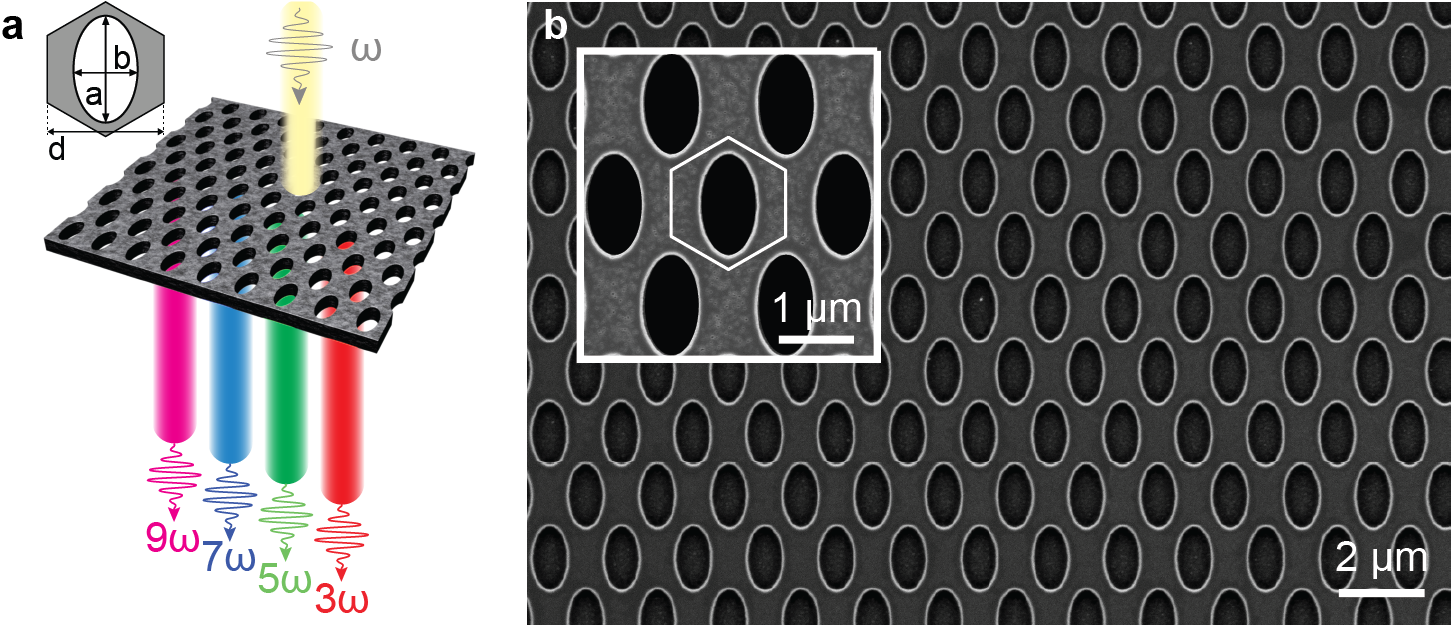}
    \caption{ {\bf Dielectric membrane metasurface for the observation of high-harmonic generation.}  (a) Concept of odd and even high-harmonic generation from resonant membrane  metasurfaces. (b) SEM image of the metasurface and the unit cell. Parameters are the following h=1 $\mu$m, d=1.69 $\mu$m, a=0.73 $\mu$m, b=0.4 $\mu$m. }
    \label{fig:concept}
\end{figure*}

Free-standing dielectric membranes provide a minimal, substrate-free platform for HHG, effectively mitigating the asymmetries introduced by underlying substrates. Recent studies have demonstrated that silicon membranes, which support Fabry-P\'erot resonances, support HHG at least up to the seventh order, with power-law exponents that align well with theoretical predictions~\cite{hallman2025high}. Furthermore, high-$Q$ free-standing membranes have been proposed for sensing applications due to their narrow resonant line-widths and exceptional mechanical stability~\cite{adi2024trapping, rosas2025enhanced}.

Herein, we consider HHG in structured free-standing membrane metasurfaces supporting ultra-high-Q resonances arising from a q-BIC mode. Such a system uniquely integrates the strong field enhancement with the simplicity inherent to a substrate-free design. Our experimental results reveal a striking enhancement in the HHG signals as compared to unpatterned membrane, exceeding three orders of magnitude for the seventh harmonic. It is also accompanied by the emergence of the ninth harmonics observable exclusively in the vicinity of resonance. Notably, our findings indicate that the harmonic power dependences deviate substantially from the classical perturbative laws, manifesting \textit{non-integer} power scaling. Theoretical simulations reveal that this phenomenon is enabled by the strong field enhancement at the q-BIC resonance, which induces a complex cross-talk between the nonlinear polarizations of different orders otherwise achievable only for significantly higher laser fluences. These results establish free-standing resonant membranes as a promising and novel platform for probing and controlling the onset of extreme regimes of high harmonic generation.



\section{Metasurface design and optical properties}

Our metasurface design is based on a free-standing silicon membrane perforated with a hexagonal grid of elliptical apertures (Fig.~\ref{fig:concept}). It is fabricated using wafer-scale compatible clean-room processes (see ``Methods'') from a $h=1~\um$ thick silicon-on-insulator base. SEM image of the structure is shown in Figure~\ref{fig:concept}b, with an inset showing a zoomed-in area with a marked unit cell.

\begin{figure}[hb]
    \centering
    \includegraphics[width=0.99\linewidth]{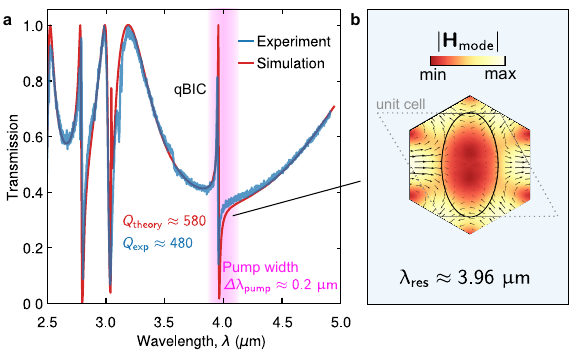}
    \caption{ {\bf Linear properties of resonant metasurfaces}. (a) Theoretical and experimental linear transmission spectra of the metasurface. (b) Magnetic field distribution in the unit cell at the quasi-BIC (qBIC) resonance.}
    \label{fig:linear_theory}
\end{figure}

\begin{figure*}
    \centering
    \includegraphics[width=0.9\linewidth]{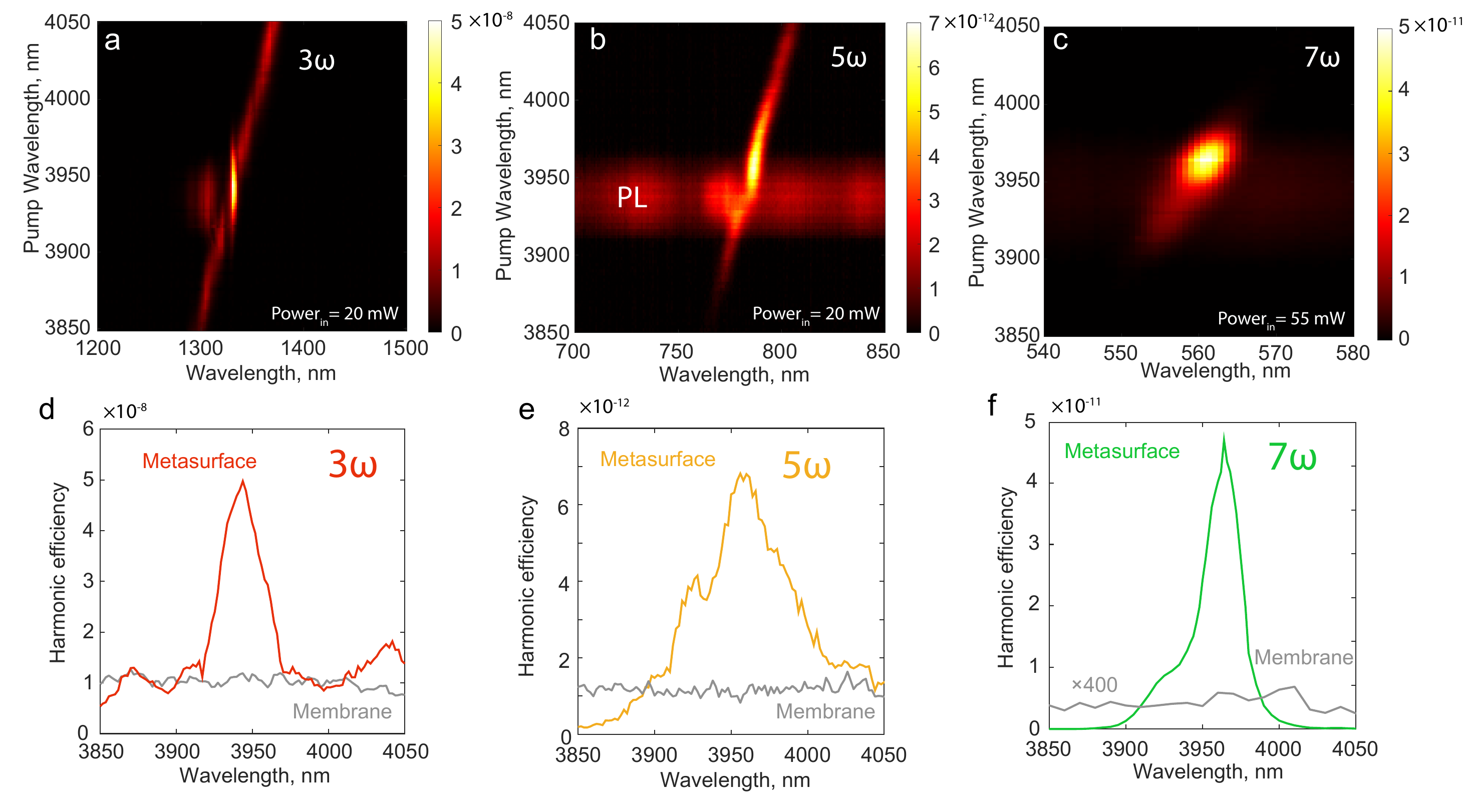}
    \caption{{\bf Experimental results.} Examples of the resonant high-harmonic generation. Generation spectra for (a) third, (b) fifth, and (c) seventh harmonics pumped at the various wavelengths. The third and fifth harmonics are pumped at $20$~mW, whereas the seventh harmonic is pumped at $55$~mW. Efficiency of (d) third, (e) fifth and (f) seventh harmonics for the metasurface and unpatterned membrane, for various pump wavelengths.}
    \label{fig:resonant}
\end{figure*}

Our design is based on symmetry-protected BIC mode that is supported by a membrane with a regular hexagonal lattice of circular apertures\cite{overvig2020selection,shakirova2025molecular}. This BIC mode can then be transformed into a q-BIC upon changing of the ellipticity of the aperture, which leads to manifestation of a narrow transmission resonance. We optimize the structure parameters to match the resonance frequency of this mode with the center of the tuning range of the laser used in nonlinear experiments. The resulting design features a regular hexagon unit cell with short diagonal $d = 1.69~\um$, an elliptical aperture with axes of $a = 0.74~\um$ and $b = 0.4~\um$, oriented along the main diagonal of the unit cell (Fig.~\ref{fig:concept}a,b). 
Figure~\ref{fig:linear_theory}a shows the calculated mid-IR transmission spectrum (red curve) that reveals several pronounced resonant features, including the Fano-shaped q-BIC resonance at $3.96~\um$. Figure~\ref{fig:linear_theory}b illustrates the electric and magnetic field distributions within a unit cell, revealing significant electromagnetic field localisation, a critical condition for achieving efficient high harmonic generation.
The experimental linear transmission spectrum, shown with a blue curve in Figure~\ref{fig:linear_theory}a, aligns exceptionally well with the theoretical predictions (see Methods for further experimental details). This close agreement not only validates our numerical models but also attests to the high quality of the fabricated metasurfaces, as evidenced by a simulated Q-factor of 580 for the quasi-BIC versus an experimentally measured value of 480.

\section{High-harmonic generation}
\subsection{Experimental results}
In the nonlinear experiments, we pump the metasurface with a tunable femtosecond laser operating in the \SIrange{3.85}{4.05}{\um} window and capture the harmonic signals in transmission both in near-infrared and visible spectral ranges (see also ``Methods'' section). In Figure~\ref{fig:resonant}a, the third harmonic generation (THG) spectrum is plotted as a function of the pump wavelength at a fixed pump power of 20~mW, which corresponds to a fluence of $\approx$2~mJ$\cdot\text{cm}^{-2}$. When the pump wavelength is detuned from the q-BIC resonance, THG spectrum shows a broad peak defined by the pump pulse linewidth which experiences an respective linear spectral shift. In stark contrast, when the pump wavelength fits within a band of $\approx$50~nm near the qBIC resonance, a narrow peak emerges from the broad harmonic spectrum. This peak aligns precisely with the tripled frequency of the qBIC mode and shows considerably stronger HHG signal. The red curve in Figure~\ref{fig:resonant}d tracks the maximum of the recorded THG intensity for each pump wavelength, revealing that the off-resonance THG conversion efficiency remains constant at approximately \num{e-8}, while the resonant excitation amplifies the efficiency up to fivefold. For comparison, the same measurements were performed on an unpatterned membrane (grey curve in Figure~\ref{fig:resonant}d), where THG remained constant within the pump laser tuning range.

\begin{figure*}
    \centering
    \includegraphics[width=0.9\linewidth]{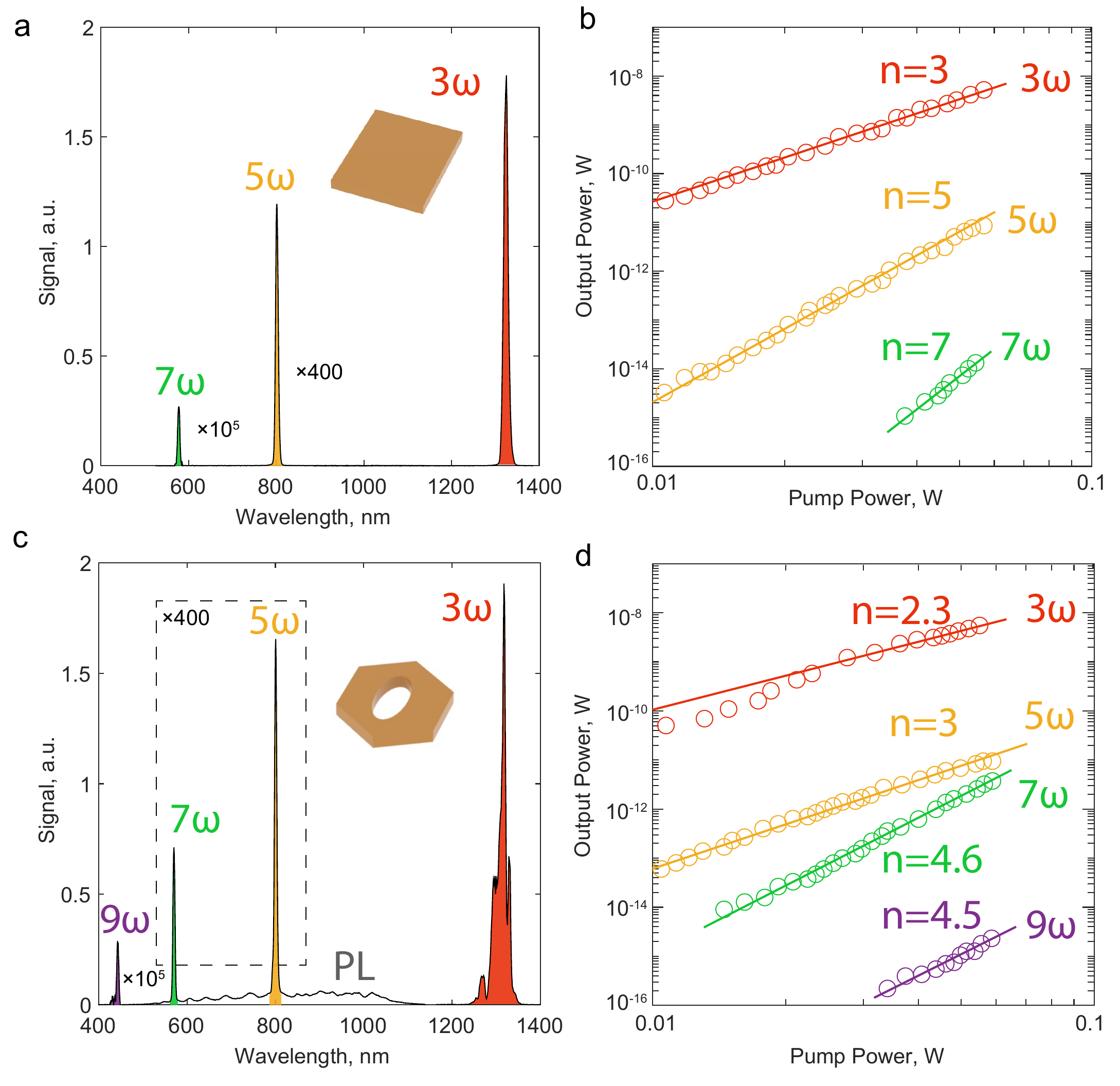}
    \caption{{\bf Experimental results.} High-harmonic generation from the membrane metasurface and unpatterned free-standing membrane. (a) Spectra and (b) power dependencies of harmonics from the unpatterned membrane. Spectra for a pump power of $50$~mW. The pump wavelength is 3.95~µm for the third harmonic and 3.96~µm for the fifth and seventh harmonics. (c) Spectra and (d) power dependencies of harmonics from the resonant membrane.}
    \label{fig:power1}
\end{figure*}

We further expand our study to higher odd harmonics. Figure~\ref{fig:resonant}b presents the dependence of the fifth-harmonic intensity on the pump wavelength at a constant pump power of 20~mW. Consistent with the observations for THG, the fifth harmonic remains relatively uniform away from resonance, with an enhancement in the vicinity of the quasi-BIC resonance.
The dependences of peak fifth harmonic conversion efficiencies on the pump wavelength presented in Figure~\ref{fig:resonant}e show that the metasurface (orange curve) provides over 7-fold enhancement over the unpatterned membrane (grey curve) near the q-BIC resonance. Notably, in addition to the resonant harmonic signal enhancement, the measured spectra reveal broad background emission when the pump wavelength matches the q-BIC resonance. We attribute this background  to silicon photoluminescence, which is enhanced by the resonant field localization similarly to the harmonics.

\begin{figure*}
    \centering
    \includegraphics[width=0.9\linewidth]{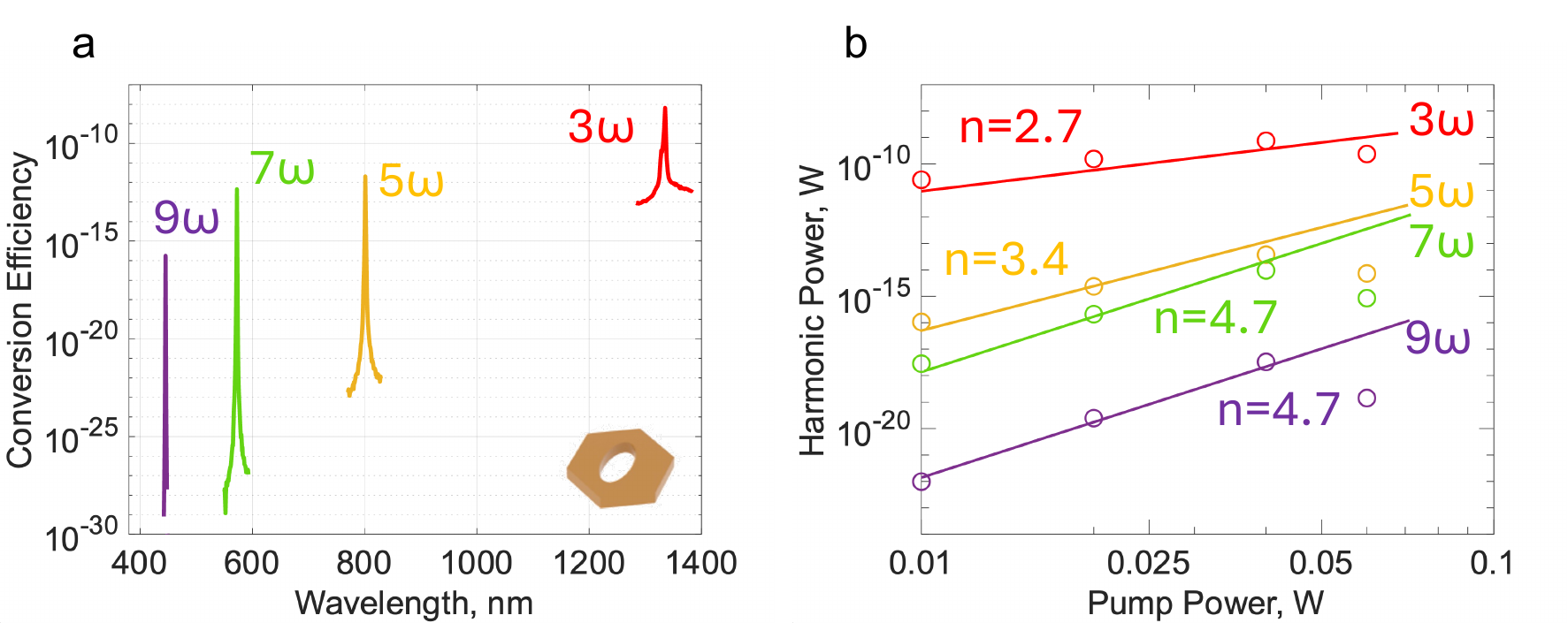}
    \caption{{\bf Theoretical results.} Calculated high-harmonic generation from the resonant membrane metasurfaces. (a) Spectra for a pump power of $50$~mW and (b) power dependencies of generated harmonics for variable input powers.}
    \label{fig:simulations1}
\end{figure*}

Finally, Figure~\ref{fig:resonant}c shows the spectra of seventh harmonic collected at a fixed pump power of 55~mW. They reveal that it manifests exclusively in the vicinity of the resonance. The extracted spectrum of conversion efficiency, plotted in Figure~\ref{fig:resonant}f (green curve), indicate that under resonant conditions it reaches \num{4.7e-11}, which is even higher than for fifth harmonic due to higher pump power used. This brings the enhancement over the unpatterned membrane (grey curve in Fig.~\ref{fig:resonant}f) to more than three orders of magnitude.

To reveal the underlying mechanisms of HHG signal enhancement, we study the power dependence of harmonic generation in both the metasurface and the unpatterned membrane. Figure~\ref{fig:power1}a displays together the spectra of the third, fifth, and seventh harmonics generated from the unpatterned membrane when pumped at a wavelength of 4.96~µm at an average power of 55~mW. For clarity, the signals of fifth and seventh harmonics are multiplied by scaling factors of 400 and  \num{e5}, respectively, to facilitate direct comparison. The corresponding power dependencies for each harmonic are plotted in Figure~\ref{fig:power1}b and show excellent agreement with the expected perturbative power-law scaling, with slopes perfectly matching the harmonic orders.

The harmonic spectra recorded from the metasurface at the maximum pump power (55~mW) are presented in Figure~\ref{fig:power1}c. One can immediately notice that the metasurface-driven enhancement improves drastically for higher harmonics, bringing the efficiency of 7th harmonic on par with the 5th (the intensity multipliers are added for the visual clarity similarly to Figure~\ref{fig:power1}a). Furthermore, we were able to detect the ninth harmonic signal from the metasurface. Its conversion efficiency was comparable to the one of the seventh harmonic from unpatterned membrane, which in turn did not show any 9th harmonic signal detectable with our equipment.

Figure~\ref{fig:power1}d illustrates the power dependencies of the harmonic signals obtained from the metasurface. In stark contrast to the unpatterned membrane, for which the harmonics adhere to conventional scaling laws, the metasurface exhibits a modified, \textit{non-integer} power scaling. Specifically, the third harmonic demonstrates a best-fit slope of 2.3 at higher pump powers, accompanied by an inflection at lower powers, while the fifth harmonic follows an effective slope of 3. In addition, the seventh and ninth harmonics are characterized by slopes of approximately 4.6 and 4.5, respectively. This unconventional behaviour is directly connected to the excitation of the qBIC mode, as we observe the same integer power scaling for the off-resonance excitation of the patterned membrane.  Furthermore, our experimental results indicate that efficient coupling of the pump laser to the qBIC mode, defined by the relative linewidth of the laser pulse and the resonance, is essential for the observation of enhanced HHG. In particular, we observe that for metasurfaces with Q-factors exceeding $\sim$~1000, efficient coupling to the qBIC mode is not achieved, and the resulting harmonic signals become indistinguishable from those generated by the unpatterned membrane even for resonant conditions.

 \subsection{Theoretical results}
To elucidate the underlying physics driving the unconventional HHG trends, we performed rigorous numerical simulations by solving the nonlinear wave equation in the frequency domain. In the simulation, we accounted for the spatial field distribution within the metasurface and incorporated all pertinent nonlinear source terms derived from a perturbative expansion of the nonlinear polarization (further details on the numerical results are available in Section~\ref{sec:Methods}). The calculated conversion efficiencies for transmitted third, fifth, seventh, and ninth harmonic signals are shown in Figure~\ref{fig:simulations1}a, assuming a fixed 50~mW pump power and pump wavelength matching the qBIC resonance (3.96~$\mu$m).  The peak conversion efficiencies were estimated at \num{6.4e-9} for third harmonic, \num{2.1e-12} for fifth harmonic, \num{4.4e-13} for seventh harmonic and \num{1.7e-16} for ninth harmonic, in good agreement with the experimental data.

We then numerically estimated the output power scaling as a function of the input pump power. As illustrated in Figure~\ref{fig:simulations1}b, our simulations reveal that even under these resonant conditions, a perturbative framework can accurately capture the observed trends. It is important to note that reproducing these trends necessitated the inclusion of nonlinear susceptibilities up to the ninth order, as well as the feedback of the harmonics at the pump frequency, incorporating both self- and cross-phase modulation effects. 
These results highlight that the observed non-integer power scaling is a direct consequence of strong field enhancement at the qBIC resonance that boosts the contribution of higher order nonlinear susceptibilities, thus modifying the effective nonlinearity of the system.

\section{Conclusions}

We have studied the high-harmonic generation from a free-standing silicon membrane metasurfaces supporting qBIC resonances. We demonstrate resonant enhancement of HHG signal when the pump matches the qBIC wavelength. The enhancement factor grows with the order of the nonlinear process involved and allows the observation of 9th harmonic that is beyond the detection limit for an unpatterned membrane.

We also reveal for the first time the deviation from the conventional power scaling of HHG driven by the qBIC resonance. In stark contrast to the unpatterned membrane, for which the power laws strictly adhere to the order of the nonlinear process, for resonant qBIC metasurfaces we observe \textit{non-integer} power dependencies. Our numerical analysis further demonstrates that a complete, generalised perturbative framework that incorporates higher-order susceptibilities and accounts for the feedback effects at the pump frequency as well as cross-phase modulation between the harmonics is sufficient to capture all salient features of the highly resonant metaphotonic system. This highlights that strong field enhancement enabled by the physics of qBIC invokes extreme regimes of high-harmonic generation, providing deep insights into the transformative role of high-Q metasurfaces in nonlinear optical processes.

\section{Methods}\label{sec:Methods}

\subsection{Device Fabrication}

Silicon (\text{Si}) membrane metasurfaces were fabricated starting from Silicon on insulator (SOI) wafers: 1~µm Device Si - 1~µm buried oxide (SiO$_2$) - 250 µm Handle Si. First, the backside opening are defined lithographically in 3~µm SiO$_2$ (Plasma-Therm Corial D250L PECVD) using direct laser writing (Heidelberg Instruments Maskless Aligner MLA 150, AZ ECI 3027, 3000~rpm) and fluorine based deep reactive ion etching (DRIE) (SPTS Advanced Plasma System). Second, the device Si is patterned by a single step electron beam lithography (Raith EBPG5000, PMMA 495k 4wt\% in Anisole, 4000~rpm) followed by DRIE (Alcatel AMS 200 SE). Third, the membranes are opened by etching the handle silicon through the previously defined SiO2 mask using a DRIE Bosch process (Alcatel AMS 200 SE). Last, the buried oxide layer is removed from under the membranes by Hydrofluoric acid (HF) vapor etching (SPTS µEtch).

\subsection{Infrared spectroscopy}

We obtained the infrared (IR) transmission spectra at normal incidence using a Bruker Vertex 80v FT-spectrometer with an IR Microscope attachment (HYPERION 3000) equipped with a liquid nitrogen cooled MCT detector. The metasurfaces are excited using a ZnSe lens with the focal length of 50~mm mildly focusing linearly polarized IR light on the sample surface. Transmitted light was collected with another 25~mm lens equipped with an additional iris placed at its back focal plane. Closing the iris allowed for limiting the numerical aperture of the system down to approximately 0.02 and thus suppressing the unwanted signal from oblique excitation angles. Signal collection area is restricted to an approximately 300~µm square central region of the membrane by a double-blade aperture placed in the conjugate image plane of the IR microscope. The sample chamber was constantly purged with dry air to provide a stable low level of humidity.

\subsection{Nonlinear spectroscopy}

The sample was pumped in the mid-IR range from 3.85~µm to 4.05~µm. The laser system consists of 1030~nm laser (Ekspla Femtolux 3) and an optical parametric amplifier (MIROPA from Hotlight Systems). The laser has a pulse duration of 250~fs and a repetition rate of 5.14~MHz. The optical parameter amplifier produces mid-IR radiation as idler pulses from the amplification of continuous wave spectrally narrow seed lasers in the near-IR spectrum. The mid-IR radiation was focused with the CaF$_2$ lens with a focus of 40~mm on the sample. The laser spot had a diameter of about 30~µm. The harmonic signal was collected by Mitutoyo Plan Apo NIR Infinity Corrected Objective X20 NA=0.4 microscope objective and detected with a Peltier-cooled spectrometer Ocean Optics QE Pro for the visible range and Ocean Optics NIR Quest for near-IR. The setup diagram can be found in Figure~S4. To measure the conversion efficiency of the generated harmonics, we used an Ophir PD300-IR power meter and spectrometers. The power of the third harmonic was measured directly with the power meter, while its spectrum was recorded by the spectrometer. This allowed us to calibrate the spectrometer counts against absolute power, particularly in cases where the signal was below the detection threshold of the power meter. Using this calibration, along with the wavelength-dependent quantum efficiency of the spectrometer, we estimated the conversion efficiencies of the higher-order harmonics based on the spectrometer measurements.

\subsection{Numerical Modelling}

We performed our numerical simulations using the finite element method implemented in COMSOL Multiphysics. In our approach, we solve the nonlinear wave equation in the frequency domain, rigorously accounting for the spatial field distribution within the metasurface and incorporating all pertinent nonlinear source terms derived from a perturbative expansion of the nonlinear polarization. The silicon bulk nonlinear susceptibilities for the third, fifth, seventh, and ninth orders are assumed to be isotropic and are integrated with a dispersion profile extracted via an experimental-numerical ellipsometry procedure\cite{rodriguez2021retrieving}. This dispersion is subsequently extended to higher orders using an atomic field scaling approach \cite{hallman2025high, boyd2008nonlinear, hallman2023harmonic, rodriguez2021retrieving}.
In our implementation, the electric field is expanded as a vector sum over its harmonic components; however, we retain only those nonlinear terms that depend on powers of the fundamental pump field, deliberately neglecting contributions arising solely from higher harmonics. This assumption is well justified by the experimentally observed low power levels at harmonic frequencies. Conversely, owing to the strong field localisation in the silicon metasurface, we preserve all nonlinear terms that contribute to pump depletion, self-phase modulation, cross-phase modulation among the harmonics, and higher-order effects at each harmonic frequency driven by the pump.

\section*{Acknowledgements} 
This work was supported by the Australian Research Council (Grant No. DP210101292) and the International Technology Center Indo-Pacific (ITC IPAC) via Army Research Office (contract FA520923C0023). The EPFL authors thank the Swiss State Secretariat for Education, Research and Innovation (SERI) for a financial support under the contract numbers of 22.00018 and 22.00081 in connection to the projects from European Union’s Horizon Europe Research and Innovation Programme under agreements 101046424 (TwistedNano) and 101070700 (MIRAQLS). M.A.V. acknowledges a financial support from NATO Science for Peace and Security program (Grant no. 5984)

\bibliography{refs}
	
\end{document}